\documentclass[11pt,a4paper]{emulateapj}
\usepackage{epsfig}
\usepackage{amsmath}
\usepackage{natbib}
\usepackage{color}
\usepackage{wasysym}

\begin{document}
\title{The Evolution of Stellar Rotation and the hydrogen atmospheres of habitable-zone Terrestrial Planets}
\author{C.~P.~Johnstone$^{1}$, M.~G\"{u}del$^{1}$, A.~St\"{o}kl$^{1}$, H.~Lammer$^{2}$, L.~Tu$^{1}$, K.~G.~Kislyakova$^{2}$, T.~L\"{u}ftinger$^{1}$, P.~Odert$^{2}$, N.~V.~Erkaev$^{3,4}$, E.~A.~Dorfi$^{1}$ }
\affil{
$^{1}$University of Vienna, Department of Astrophysics, Vienna, Austria\\
$^{2}$Space Research Institute, Austrian Academy of Sciences, Graz, Austria\\
$^{3}$Institute for Computational Modelling, Siberian Division of Russian Academy of Sciences, Krasnoyarsk, Russian Federation \\ 
$^{4}$Siberian Federal University, Krasnoyarsk, Russian Federation
}
\date{\today}

\begin{abstract}
Terrestrial planets formed within gaseous protoplanetary disks can accumulate significant hydrogen envelopes. 
The evolution of such an atmosphere due to XUV driven evaporation depends on the activity evolution of the host star, which itself depends sensitively on its rotational evolution, and therefore on its initial rotation rate.  
In this letter, we derive an easily applicable method for calculating planetary atmosphere evaporation that combines models for a hydrostatic lower atmosphere and a hydrodynamic upper atmosphere. 
We show that the initial rotation rate of the central star is of critical importance for the evolution of planetary atmospheres and can determine if a planet keeps or loses its primordial hydrogen envelope. 
Our results highlight the need for a detailed treatment of stellar activity evolution when studying the evolution of planetary atmospheres. 
\end{abstract}

\section{Introduction}

Planetary atmospheres evolve under the influence of the winds and high energy radiation of their hosts stars. 
High-energy radiation drives chemistry (\citealt{2008JGRE..113.5008T}; \citealt{2013Icar..226.1678K}; \citealt{2014ApJ...795..132S}; \citealt{2014AA...571A..94S}; \citealt{2015Icar..250..357C}) and heating, causing expansion and mass loss (\citealt{2005ApJ...621.1049T}; \citealt{2013AsBio..13.1011E}; \citealt{2015AsBio..15...57L}).
In addition, atmospheres exposed to stellar winds often experience significant additional loss (\citealt{2007AsBio...7..185L}; \citealt{2010Icar..210....1L}; \citealt{2011SSRv..162..309L}; \citealt{2013AsBio..13.1030K}).

Atmospheric evolution is fundamentally linked to the evolution of stellar winds and XUV emission (where XUV is X-ray+EUV).
A star's XUV emission originates from magnetically heated chromospheric and coronal plasma (\citealt{2004AARv..12...71G}; \citealt{2006MNRAS.367..917J}) and is determined primarily by the star's rotation, with rapid rotators being more active than slow rotators, except at rapid rotation where the activity saturates (\citealt{2011ApJ...743...48W}).
As a star ages, its activity declines due to spin-down (\citealt{1997ApJ...483..947G}; \citealt{2014MNRAS.441.2361V}).
Due to the fact that a star's rotation evolves differently depending on its initial rotation rate, a star's activity level is \emph{not} uniquely determined by its mass and age (\citealt{2015AA...577A..28J}; \citealt{2015AA...577L...3T}). 
Solar mass stars at ages of 1 Myr have a large distribution of rotation rates, ranging from a few to a few tens of times faster than the current solar rotation rate (\citealt{2002AA...396..513H}; \citealt{2014prpl.conf..433B}; \citealt{2015ApJ...799L..23M}). 
At this age, due to their internal structures, all stars lie above the saturation threshold in rotation. 
\citet{2015AA...577L...3T} found that the age at which a star falls out of saturation varies between $\sim$10~Myr and several hundred Myr depending on its initial rotation rate.

The simplest planetary atmospheres are those dominated by hydrogen collected from the circumstellar disk (\citealt{2014MNRAS.439.3225L}; \citealt{2015AA...576A..87S}).
For such an atmosphere to form, the planet must grow to a significant mass ($>$0.1~M$_\oplus$) before the disk dissipates.
Disk lifetimes are typically a few Myr (\citealt{2012ApJ...745...19K}), whereas standard planet formation theory suggests that terrestrial planet formation takes much longer (\citealt{2011ASL.....4..325L}).
However, the existence of low density terrestrial planets, such as those in the Kepler-11 system (\citealt{2011Natur.470...53L}), suggest that at least some terrestrial planets are able to form in the gas disk.
Furthermore, recent studies have indicated that a large fraction of observed terrestrial planets have low densities, suggesting that they have thick gaseous envelopes (\citealt{2014PNAS..11112655M}; \citealt{2015ApJ...801...41R}).

\begin{figure*}
\centering
\includegraphics[width=0.45\textwidth]{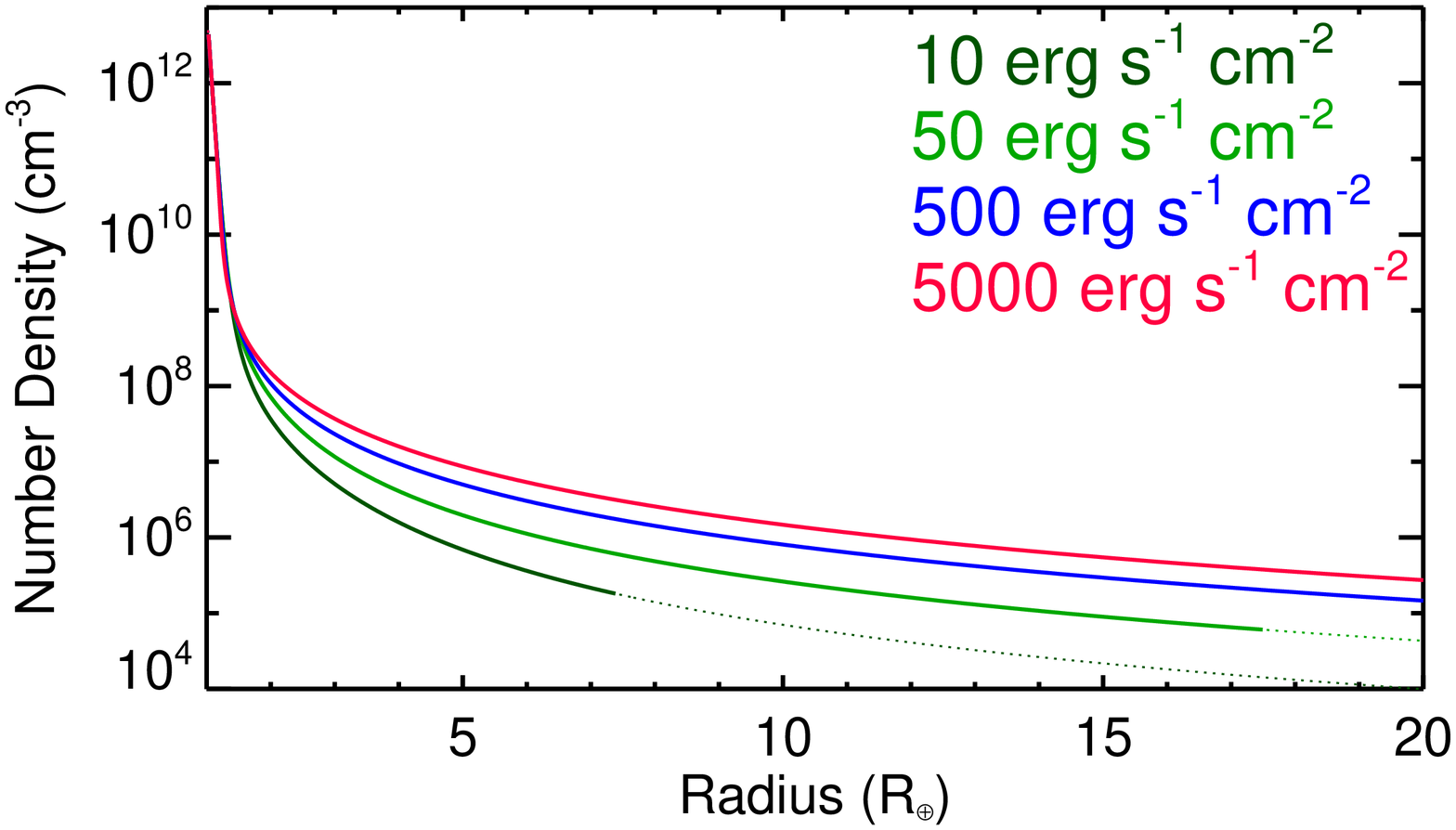}
\includegraphics[width=0.45\textwidth]{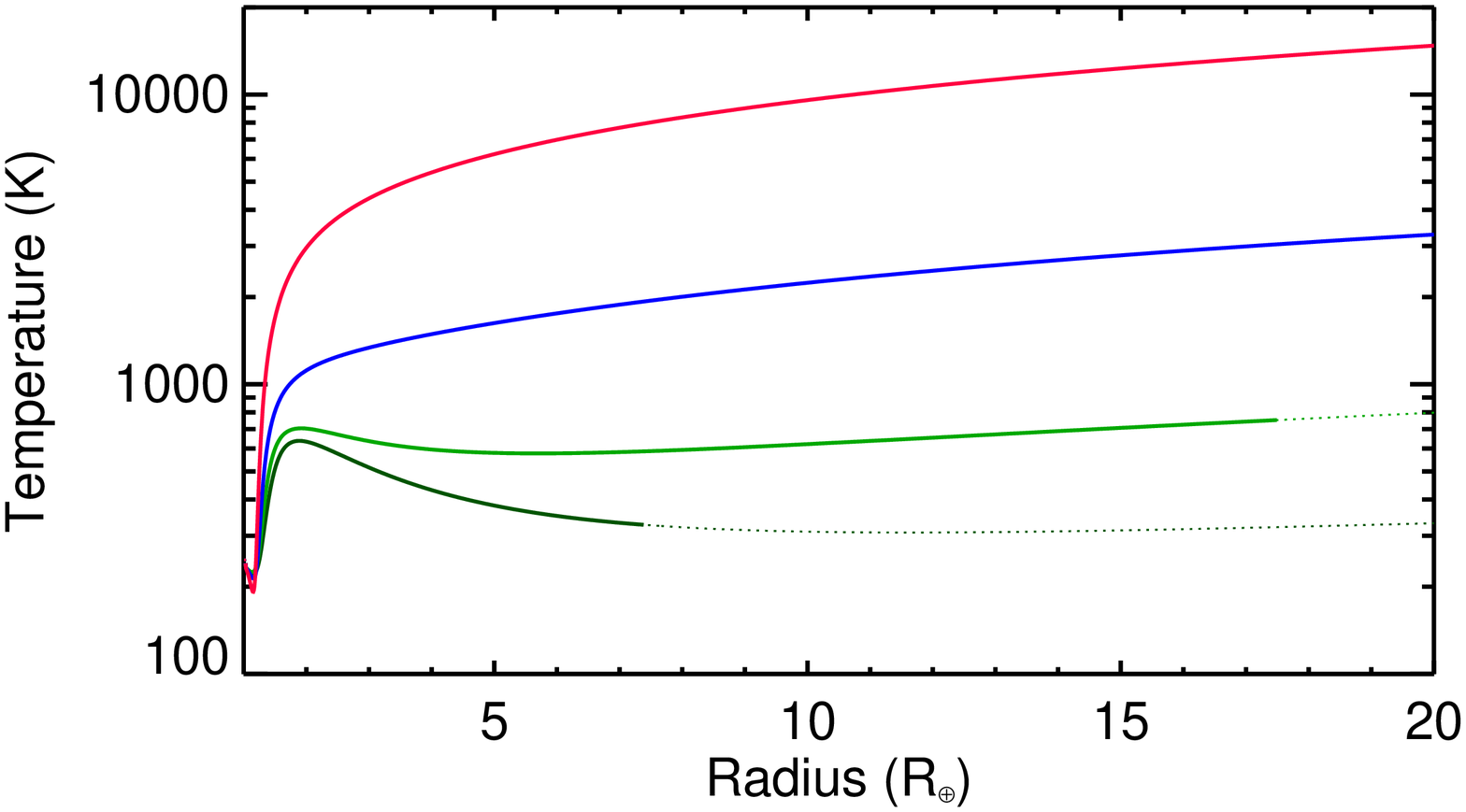}
\includegraphics[width=0.45\textwidth]{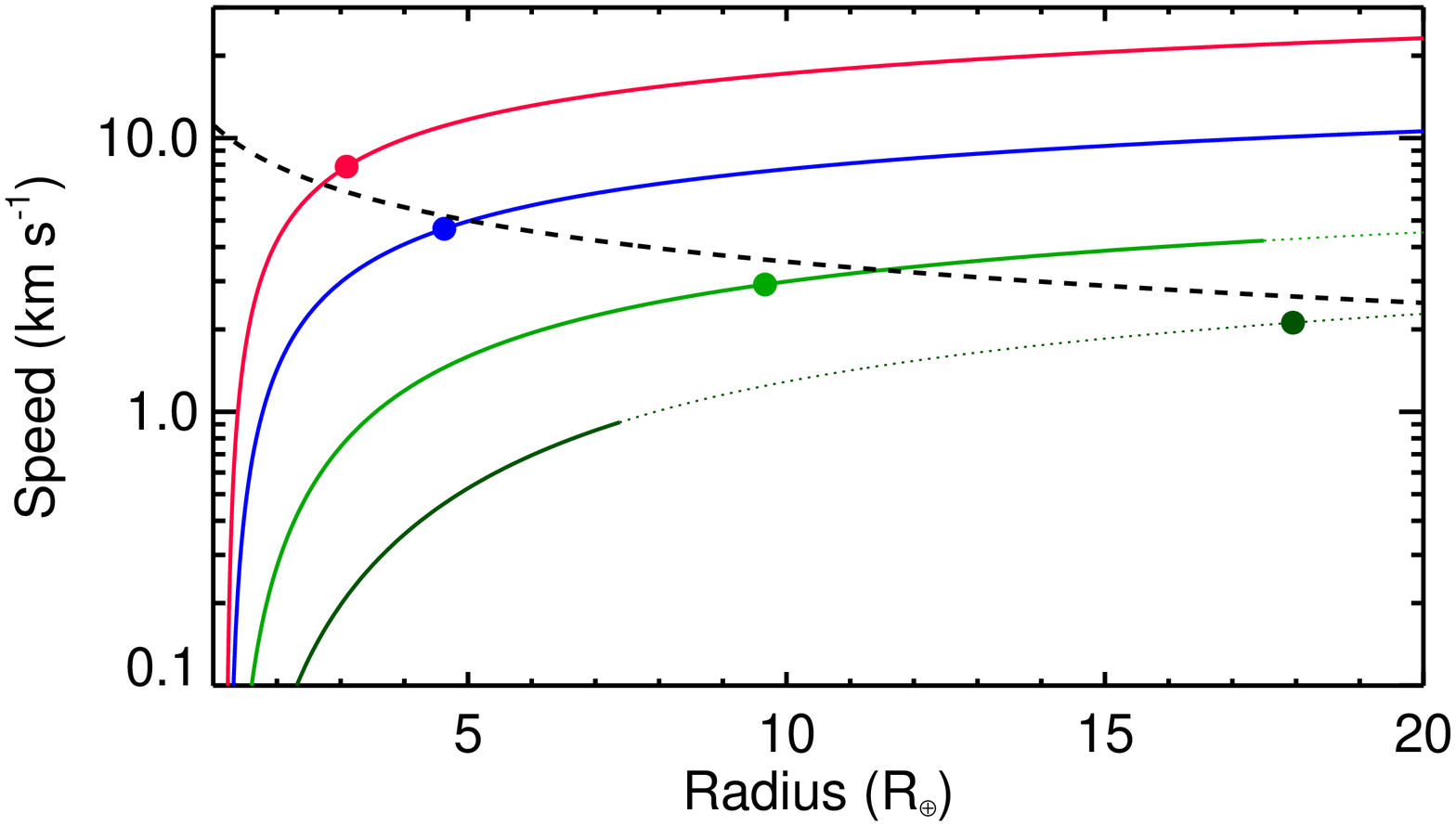}
\includegraphics[width=0.45\textwidth]{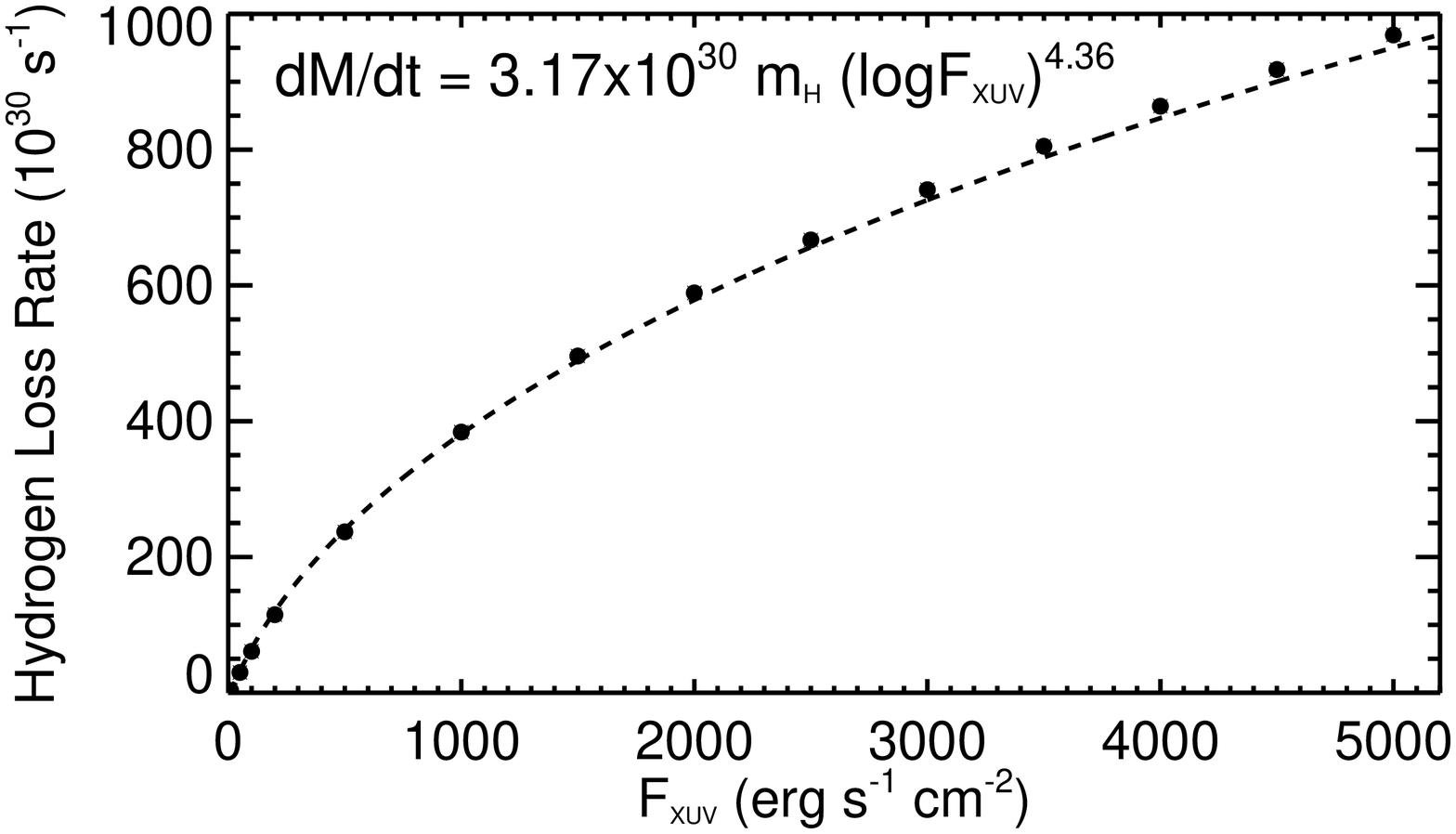}
\includegraphics[width=0.45\textwidth]{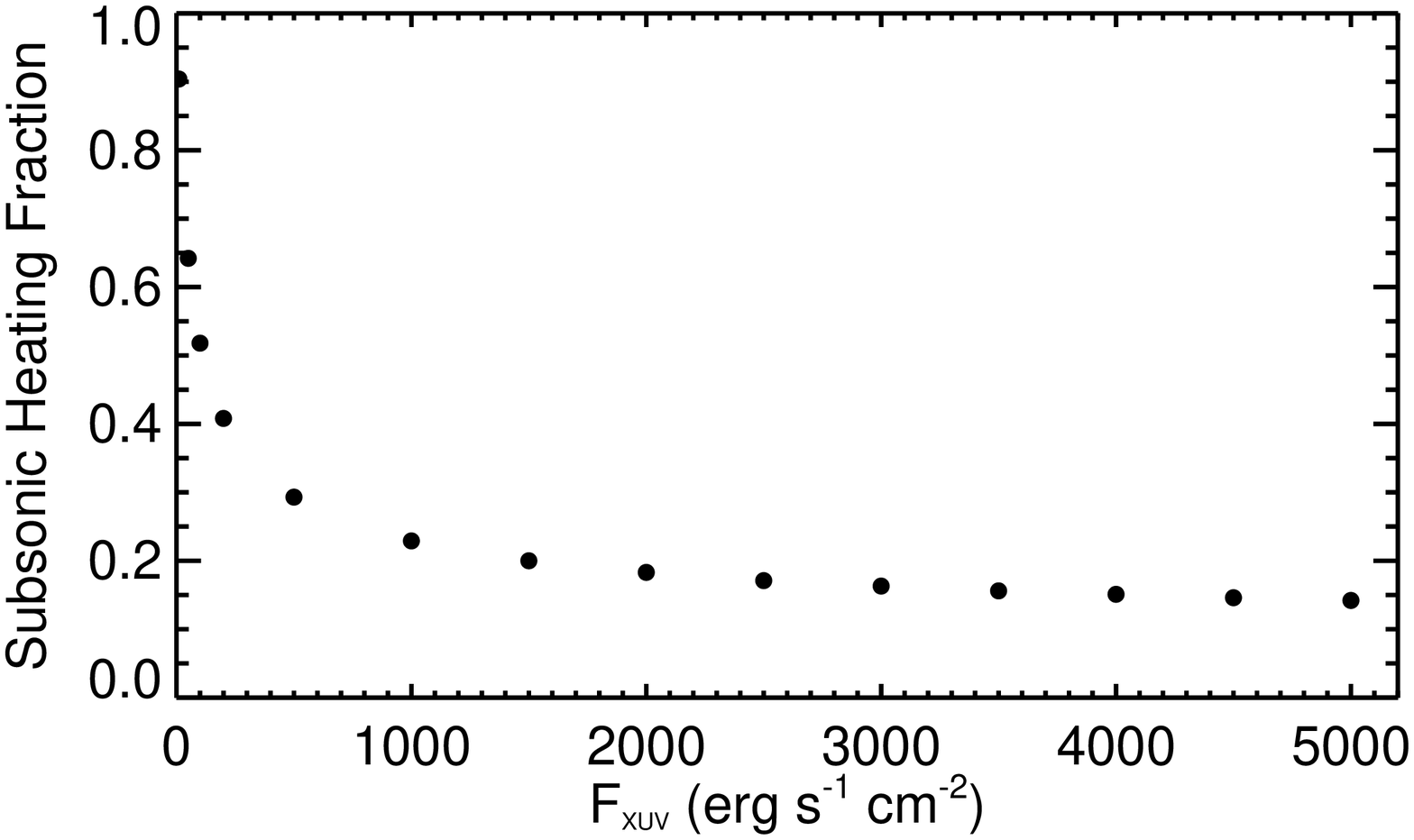}
\includegraphics[width=0.45\textwidth]{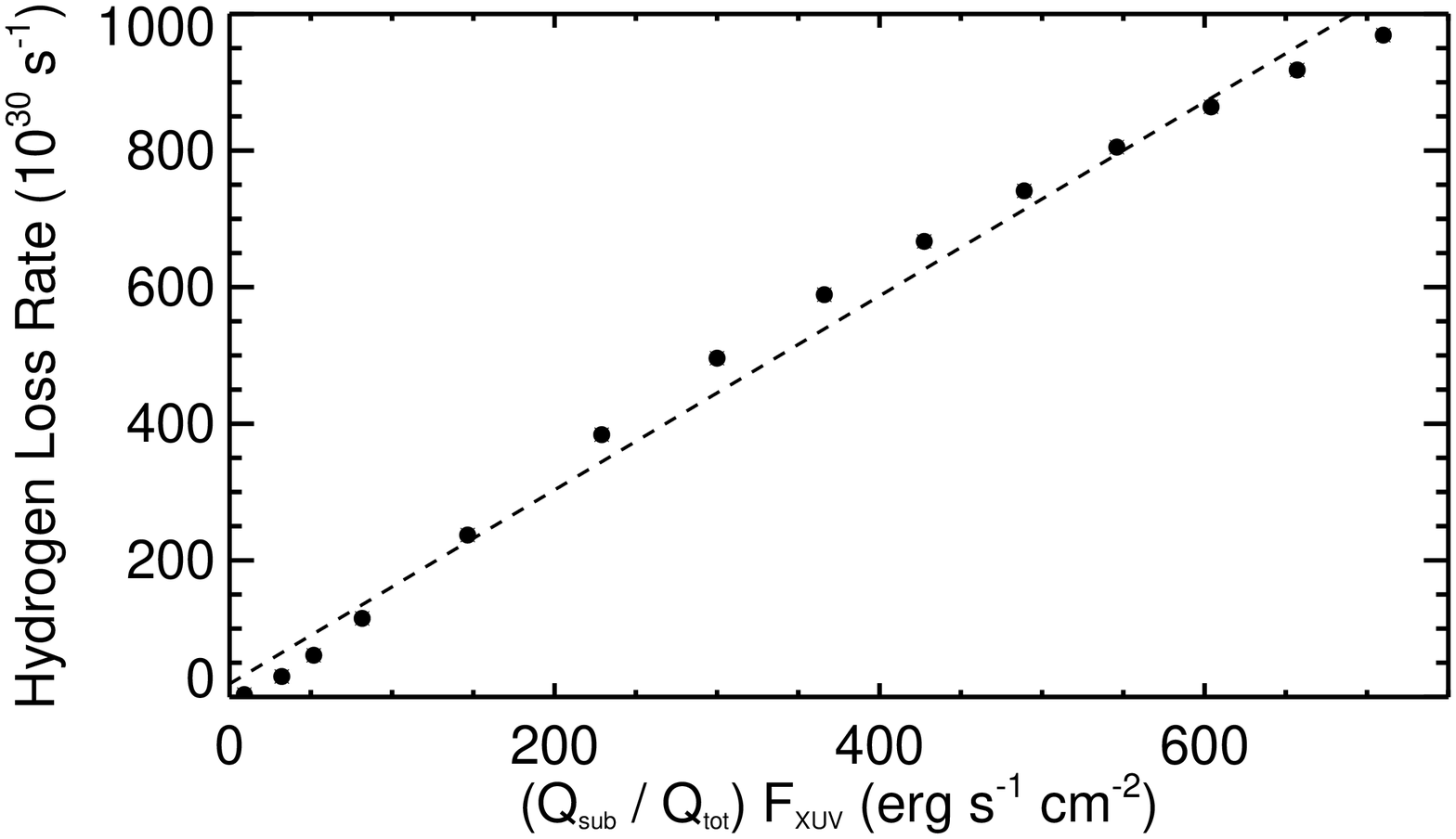}
\caption{
Hydrodynamic models of upper atmosphere expansion from an Earth mass planet with different input XUV fluxes. 
For $F_\text{XUV}$ of 10, 50, 500, and 1000~erg~s$^{-1}$~cm$^{-2}$, we show profiles of hydrogen density (\emph{upper-left}), temperature (\emph{upper-right}), and radial expansion velocity (\emph{middle-left}).
The dotted parts of the lines show the regions where the Knudsen number (defined here using Eqn.~3-5 of \citealt{2015MNRAS.448.1916E}) is greater than unity, and therefore the simulations are not reliable (the top of the solid line therefore shows the exobase).
In the velocity plot, the dashed line is the escape velocity and the circles show where the wind becomes supersonic. 
\emph{Middle-right panel}: atmospheric hydrogen loss rate as a function of  $F_\text{XUV}$. 
\emph{Lower-left panel}: subsonic heating fraction (i.e. the fraction of heat deposited below the sonic point), \mbox{$Q_\text{sub} / Q_\text{tot}$}, as a function of input $F_\text{XUV}$. 
\emph{Lower-right panel}: $\dot{M}_\text{at}$ as a function of \mbox{$(Q_\text{sub} / Q_\text{tot} ) F_\text{XUV}$}, showing that the $\dot{M}_\text{at}$ is approximately proportional to the energy deposited in the subsonic part of the wind.
} \label{fig:hydrosims}
\end{figure*}

The capture and subsequent escape of disk gas by terrestrial planets was studied by \citet{2014MNRAS.439.3225L} who modeled the first 100~Myr of the planet's life and assumed the star's activity remains saturated for the entire time.
The amount of disk gas captured is strongly dependent on the core mass, with low mass cores capturing orders of magnitude less gas than high mass cores. 
\citet{2014MNRAS.439.3225L} showed that low-mass habitable-zone terrestrial planets will not keep hydrogen envelopes for evolutionary timescales, whereas many high-mass terrestrial planets will always keep such atmospheres. 
While the planet is embedded in the disk, thermal pressure from disk gas on the atmosphere provides additional support binding it to the core.
\citet{2015AA...576A..87S} showed that when this thermal support is removed, the atmosphere can flow away at rates that also strongly depend on core mass.


In this letter, we study the importance of the initial stellar rotation rate on the XUV driven evolution of hydrogen atmospheres. 
In Section~\ref{sect:models}, we describe our model for atmospheric evolution; in Section~\ref{sect:results}, we calculate models for a range of cases; in Section~\ref{sect:discussion}, we discuss our results.

\begin{figure}
\centering
\includegraphics[trim=15mm 0mm 0mm 0mm, width=0.49\textwidth]{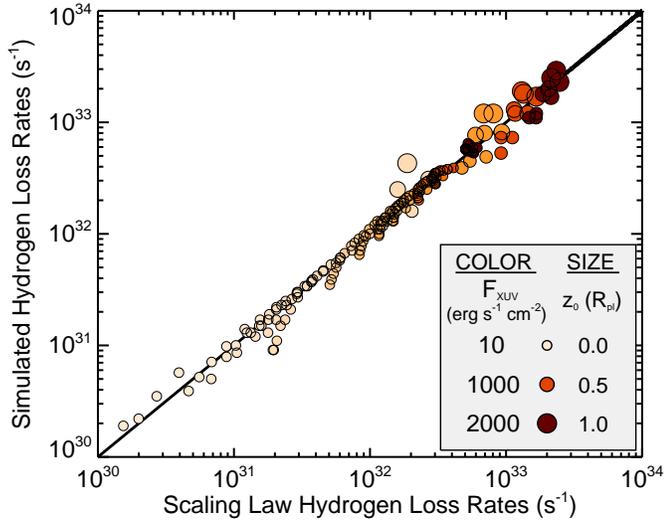}
\caption{
Simulated $\dot{M}_\text{at}$ from our hydrodynamic simulations against values from our scaling law (Eqn.~\ref{eqn:Mdoteqn}). 
Each symbol represents one simulation, with darker colours showing larger $F_\text{XUV}$ and larger symbols showing larger $z_0$.
} \label{fig:grid}
\end{figure}

\section{Atmospheric evolution model} \label{sect:models}

Our atmospheric evolution calculations combine a hydrodynamic upper atmosphere model (Section~\ref{sect:MdotModel}), a hydrostatic lower atmosphere model (Section~\ref{sect:LowerAt}), and evolutionary tracks for stellar XUV luminosity (Section~\ref{sect:TuTracks}).
In all models, we assume the planet is in the habitable zone at 1~AU around a solar mass star and has an equilibrium temperature of 250~K.
We therefore do not consider the evolutionary changes in the star's bolometric luminosity, which will be studied in future work.

\begin{figure}
\centering
\includegraphics[trim=15mm 0mm 0mm 0mm, width=0.49\textwidth]{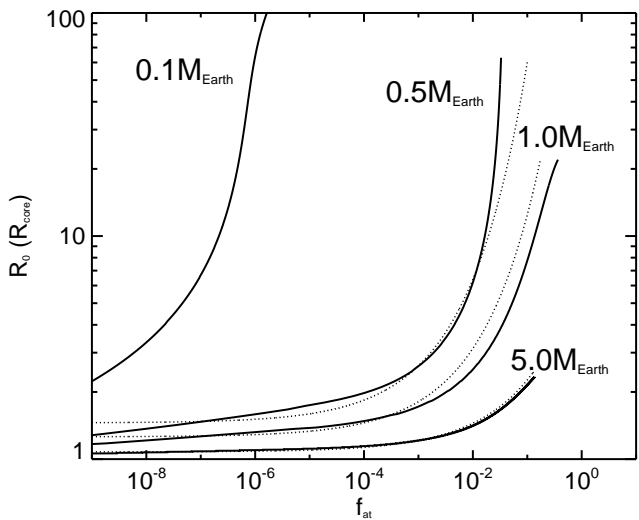}
\caption{
The value of $R_0$ (\mbox{$=R_\text{core} + z_0$}) as a function of planetary atmospheric mass, \mbox{$f_\text{at} = M_\text{at} / M_\text{pl}$}, for different planetary masses.
The dotted lines show the approximate values of $R_0$ given by Eqn.~\ref{eqn:R0scaling}. 
} \label{fig:lowerat}
\end{figure}

\subsection{Atmospheric mass loss and the $F_\text{XUV}$ dependence} \label{sect:MdotModel}

To predict the atmospheric mass loss rate, $\dot{M}_\text{at}$, we use hydrodynamic upper atmosphere simulations performed using the Versatile Advection Code (VAC; \citealt{1996ApLC..34..245T}). 
We run our simulations in 1D spherical geometry with 1000 unevenly spaced grid cells and include the planet's gravity and XUV heat deposition. 
At the base of the simulation, we assume a constant density of \mbox{$5 \times 10^{12}$~cm$^{-3}$} and a constant temperature of 250~K.
The base density was chosen so that the entire XUV flux will be absorbed within the computational domain.  
We further assume that the upper atmosphere consists purely of neutral atomic hydrogen. 
The three input parameters are the planetary mass, $M_\text{pl}$, the stellar XUV energy flux, $F_\text{XUV}$, and the radius of the base of the simulation, $R_0$ (alternatively, we often use \mbox{$z_0=R_0-R_\text{core}$}).

\begin{figure*}
\centering
\includegraphics[width=0.49\textwidth]{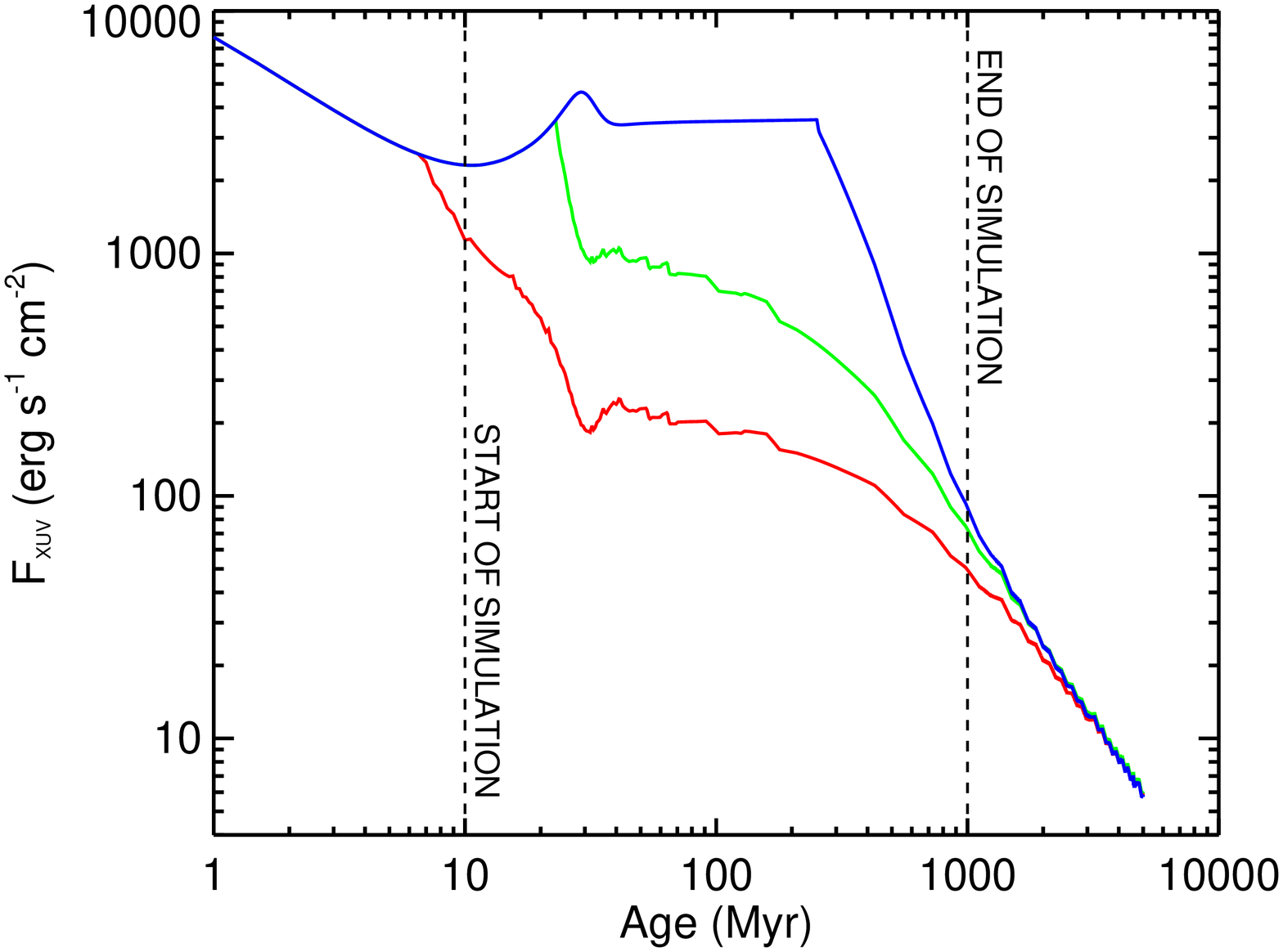}
\includegraphics[width=0.49\textwidth]{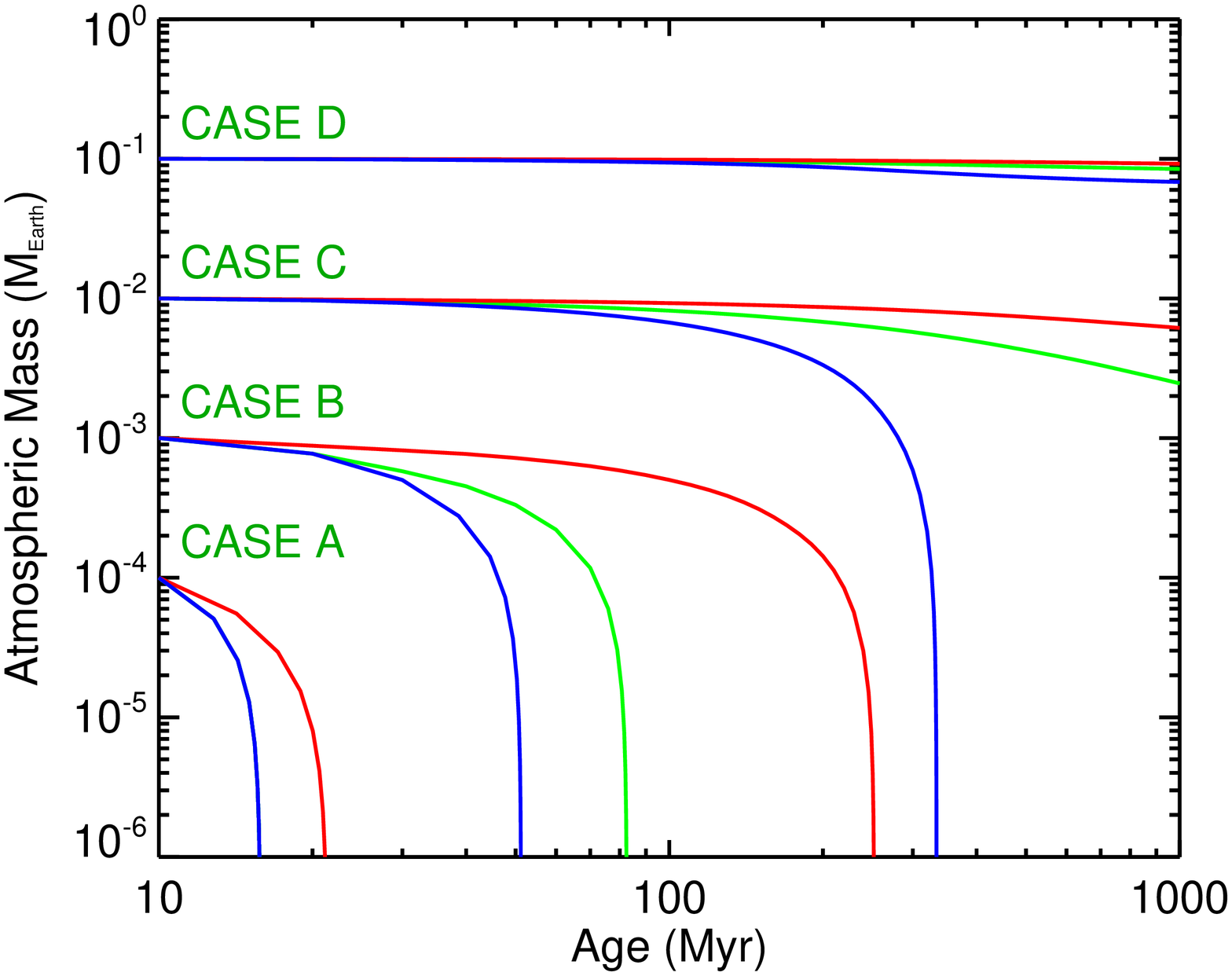}
\caption{
\emph{Left panel}: the evolution of the XUV (1-960\AA) flux received by a planet orbiting a solar mass star at 1~AU assuming separately that the star is a slow rotator (\emph{red}), an average rotator (\emph{green}), and a fast rotator (\emph{blue}). 
\emph{Right panel}: the evolution of hydrogen protoatmospheres between 10~Myr and 5~Gyr in response to the XUV irradiation for Earth mass planets. 
The four cases, marked as green symbols in Fig.~\ref{fig:AllMasses}, correspond to different initial atmospheric masses.
} \label{fig:EarthCases}
\end{figure*}

We assume that mass loss happens evenly over the entire area of the planet, excluding the shadow cast by the planet itself (i.e. over $\sim$3$\pi$~steradians). 
We average the total input flux over this area, which gives an input XUV flux at the top of the simulations of $F_\text{XUV}/3$. 
As in \citet{2009ApJ...693...23M} and \citet{2013AsBio..13.1011E}, we assume a single wavelength for all photons and use the absorption cross section from \citet{2013AsBio..13.1011E} of \mbox{$\sigma = 5 \times 10^{-18}$~cm$^{2}$}. 
We integrate the input XUV flux downwards through the atmosphere by decreasing the radiation flux by a factor of $e^{-\tau}$ when traveling through each grid cell, where \mbox{$\tau = n \sigma ds$}, $ds$ is the grid cell thickness, and $n$ is the hydrogen atom number density.
The heating of the gas within each grid cell is given by \mbox{$q = \epsilon \sigma n F_\text{XUV}$}, where $F_\text{XUV}$ is in this case the XUV flux in the grid cell and \mbox{$\epsilon = 0.15$} is the heating efficiency parameter (\citealt{2014AA...571A..94S}; \citealt{2015MNRAS.448.1916E}; \citealt{2015SoSyR..49..339I}).

To test the dependence of $\dot{M}_\text{at}$ on $F_\text{XUV}$, we run 14 models for Earth mass planets with \mbox{$z_0 = 100$~km} and $F_\text{XUV}$ ranging from 10 to 5000~erg~s$^{-1}$~cm$^{-2}$. 
The current Earth receives an $F_\text{XUV}$ of $\sim$5~erg~s$^{-1}$~cm$^{-2}$.
For very similar simulations and $F_\text{XUV}$ of 10, 50, and 100~erg~s$^{-1}$, \citet{2013AsBio..13.1011E} found hydrogen atom loss rates of \mbox{$5.0 \times 10^{30}$}, \mbox{$1.9 \times 10^{31}$}, and \mbox{$3.2 \times 10^{31}$}~s$^{-1}$. 
We find similar values of \mbox{$4.5 \times 10^{30}$}, \mbox{$3.0 \times 10^{31}$}, and \mbox{$6.1 \times 10^{31}$~s$^{-1}$}.

The results of our 14 simulations are demonstrated in Fig. \ref{fig:hydrosims}. 
Clearly, $\dot{M}_\text{at}$ does not depend linearly on $F_\text{XUV}$ because as $F_\text{XUV}$ increases, the fraction of the input energy available to lift mass away from the planet decreases. 
In our models, $\dot{M}_\text{at}$ is only influenced by heating below the sonic point. 
As $F_\text{XUV}$ increases, the sonic point moves closer to the planet leading to a smaller fraction of the energy being deposited in the subsonic region. 
A similar effect can be seen in hydrodynamic stellar wind models (e.g. see Fig. 8 and Fig. 9 of \citealt{2015AA...577A..27J}).

To derive a  scaling law for $\dot{M}_\text{at}$, we run a grid of models with a range of $M_\text{pl}$, $F_\text{XUV}$, and $z_0$. 
In total, we run 230 simulations with $M_\text{pl}$ between 0.5 and 5.0 $M_\oplus$, $F_\text{XUV}$ between 10 and 2000~erg~s$^{-1}$~cm$^{-2}$, and $z_0$ between 100~km and $1R_\text{core}$. 
Of these simulations, 46 did not give realistic results due to numerical difficulties. 
Motived by experiments with different functions, we assume that $\dot{M}_\text{at}$ is given by
\begin{equation} \label{eqn:Mdoteqn}
\dot{M}_\text{at} = a m_\text{H} M_\text{pl}^b z_0^c \left( \log F_\text{XUV} \right)^{ g \left( M_\text{pl} , z_0 \right) },
\end{equation}
where
\begin{equation}
g \left( M_\text{pl} , z_0 \right) = d M_\text{pl}^e z_0^f,
\end{equation}
$m_\text{H}$ is the mass of a hydrogen atom and \mbox{$z_0 = R_0 - R_\text{core}$} is the altitude of the base of the simulation. 
We find that \mbox{$a = 1.858 \times 10^{31}$}, \mbox{$b = -1.526$}, \mbox{$c = 0.464$}, \mbox{$d = 4.093$}, \mbox{$e = 0.249$}, and \mbox{$f = -0.022$}; in addition, $M_\text{pl}$, $z_0$, and $F_\text{XUV}$ must be in units of $M_\oplus$, $R_\oplus$, and erg~s$^{-1}$~cm$^{-2}$.
The quality of this fit to our grid of simulations is shown in Fig.~\ref{fig:grid}. 
Although our scaling law is inelegant and gives little intuitive insight into the physics of evaporation, it likely gives good estimates of $\dot{M}_\text{at}$.

\subsection{Lower atmospheric extent} \label{sect:LowerAt}

The extent of a planet's atmosphere is strongly influenced by the planet's mass, the atmospheric mass, and the atmospheric composition (\citealt{2012AA...547A.112M}; \citealt{2014ApJ...787..173H}; \citealt{2014ApJ...792....1L}). 
Based on a few simplifying assumptions, we are able to estimate $z_0$ from the planetary and atmospheric masses.
In general, in more massive atmospheres, $z_0$ is at higher altitudes, leading to higher $\dot{M}_\text{at}$.
As an atmosphere evaporates, both $z_0$ and $\dot{M}_\text{at}$ decrease; when the atmosphere is gone, \mbox{$z_0 = 0$} and Eqn.~\ref{eqn:Mdoteqn} predicts no mass loss.
 
The exact definition of $z_0$ is the altitude at which the hydrogen atom number density is \mbox{$5 \times 10^{12}$~cm$^{-3}$}, since that is what we take as the density at the base of our simulations.  
Calculating $z_0$ therefore requires the lower atmosphere density structure. 
For this, we use the initial model integrator of the TAPIR-Code (\citealt{2008AA...490.1181S}; \citealt{2015AA...576A..87S}) to solve the hydrostatic structure equations (Eqn. 4-6 of \citealt{2015AA...576A..87S}) taking into account radiative and convective energy transport.
We assume that the core density is equal to that of the Earth for all bodies, meaning that \mbox{$R_\text{core} \propto M_\text{pl}^{1/3}$}.
To describe the gas in the lower atmosphere, we use the equation of state derived by \citet{1995ApJS...99..713S} for a hydrogen and helium mixture, and the opacities for gas and dust from \citet{Freedman2008} and \citet{Semenov2003}, respectively. 
The free parameters are the flux of energy from the planetary core, $L_\text{pl}$, and the dust depletion factor, $f$. 
As in \citet{2015AA...576A..87S}, we take \mbox{$L_\text{pl} = 10^{21} \left( M_\text{pl} / M_\oplus \right)$~erg~s$^{-1}$} and \mbox{$f=0.01$}. 
In reality, these two parameters depend on the specific planet formation scenario and on the age of the system; a more detailed treatment of these parameters will be the subject of further work. 

The dependence of $R_0$ on planetary and atmospheric masses are shown in Fig.~\ref{fig:lowerat}.
For masses above $0.5M_\oplus$, this dependence is approximately described by
\begin{equation} \label{eqn:R0scaling}
\log \left( \frac{R_0}{R_\text{core}} \right) = \left( 2.5 f_\text{at}^{0.4} + 0.1 \right) \left( \frac{M_\text{pl}}{M_\oplus} \right)^{-0.7}.
\end{equation}
The dotted lines in Fig.~\ref{fig:lowerat} show the quality of this fit. 
The mass dependent upper limit for $f_\text{at}$, i.e. the value at which the lines in Fig.~\ref{fig:lowerat} turn upwards, is given by
\begin{equation} \label{eqn:fatmax}
f_\text{at,max} \approx 0.3 \left( \frac{M_\text{pl} }{ M_\oplus } \right)^{3.6}.
 \end{equation}
For a hydrogen atmosphere of a terrestrial planet, $\dot{M}_\text{at}$ can easily be estimated by combining Eqn.~\ref{eqn:Mdoteqn} with Eqn.~\ref{eqn:R0scaling}, though our scaling laws only apply to planets with equilibrium temperatures of $\sim$250~K.

\begin{figure*} 
\centering
\includegraphics[width=0.49\textwidth]{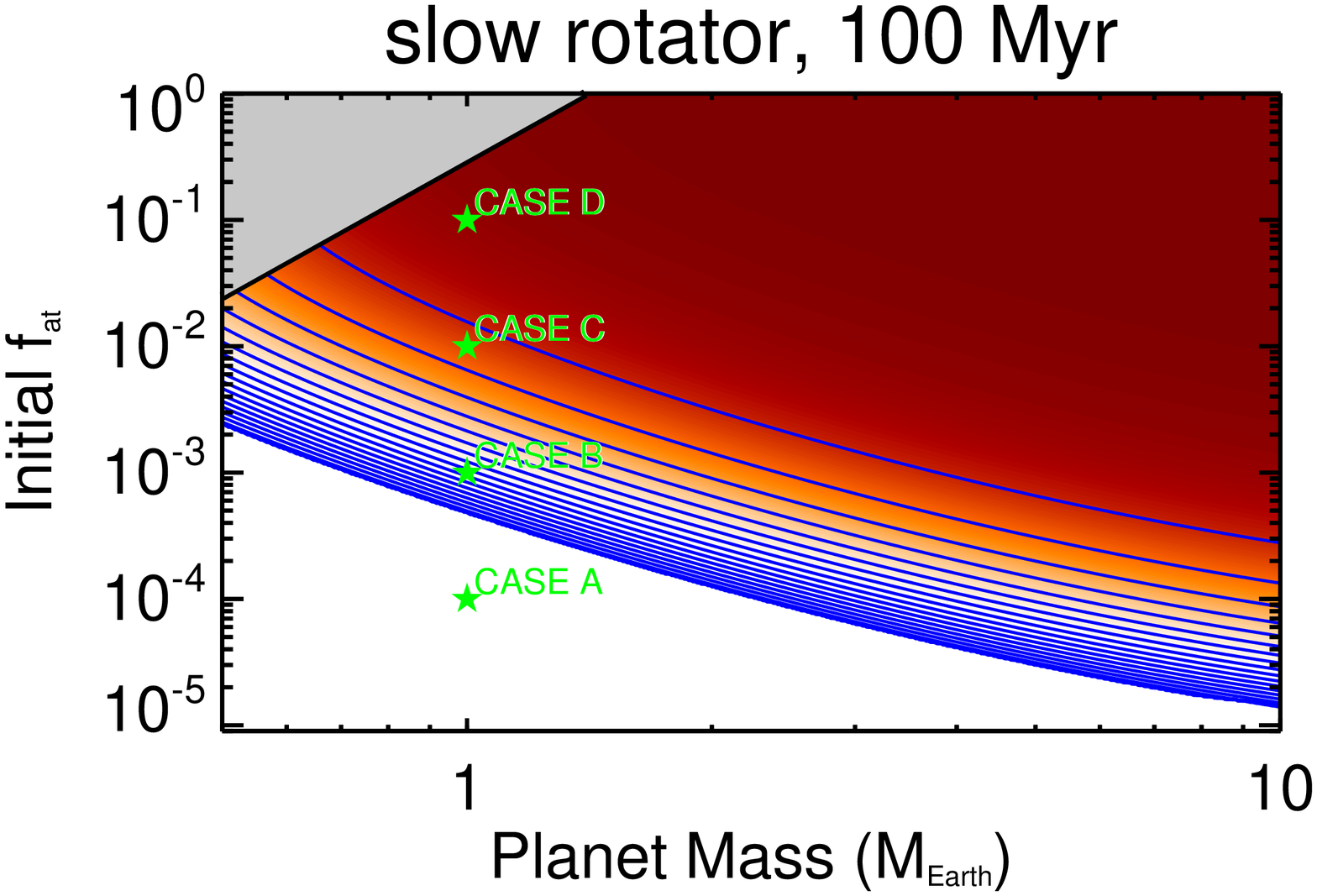}
\includegraphics[width=0.49\textwidth]{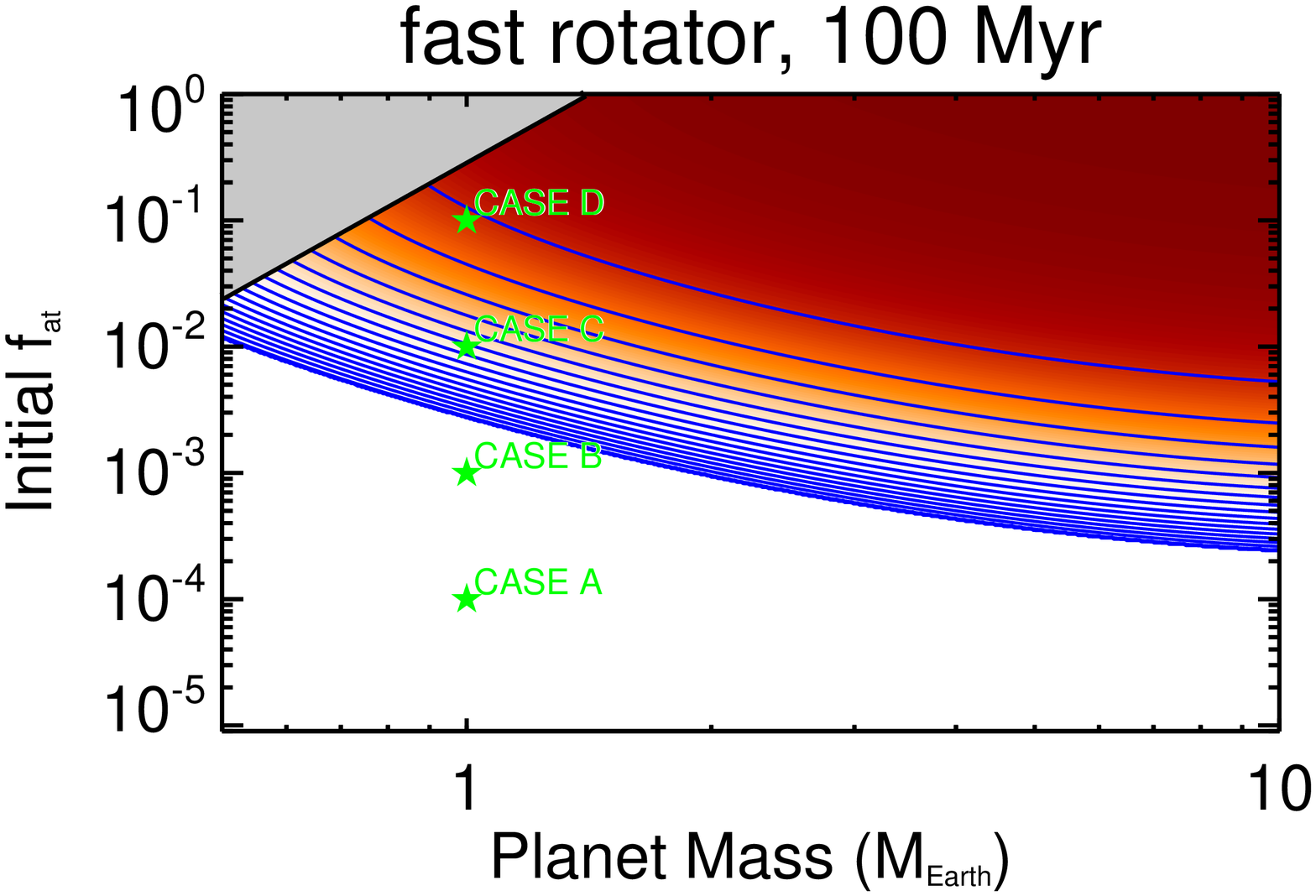}
\includegraphics[width=0.49\textwidth]{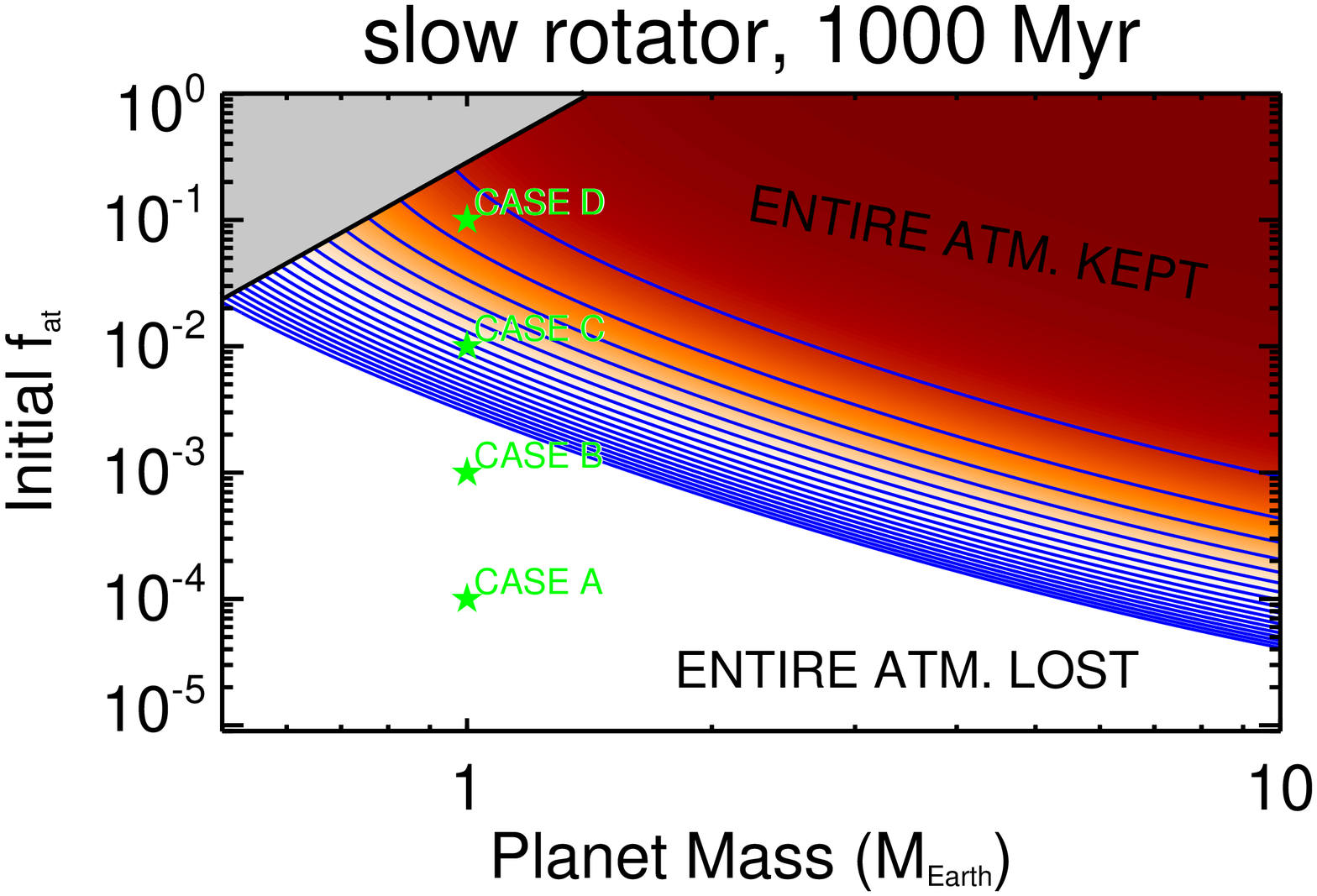}
\includegraphics[width=0.49\textwidth]{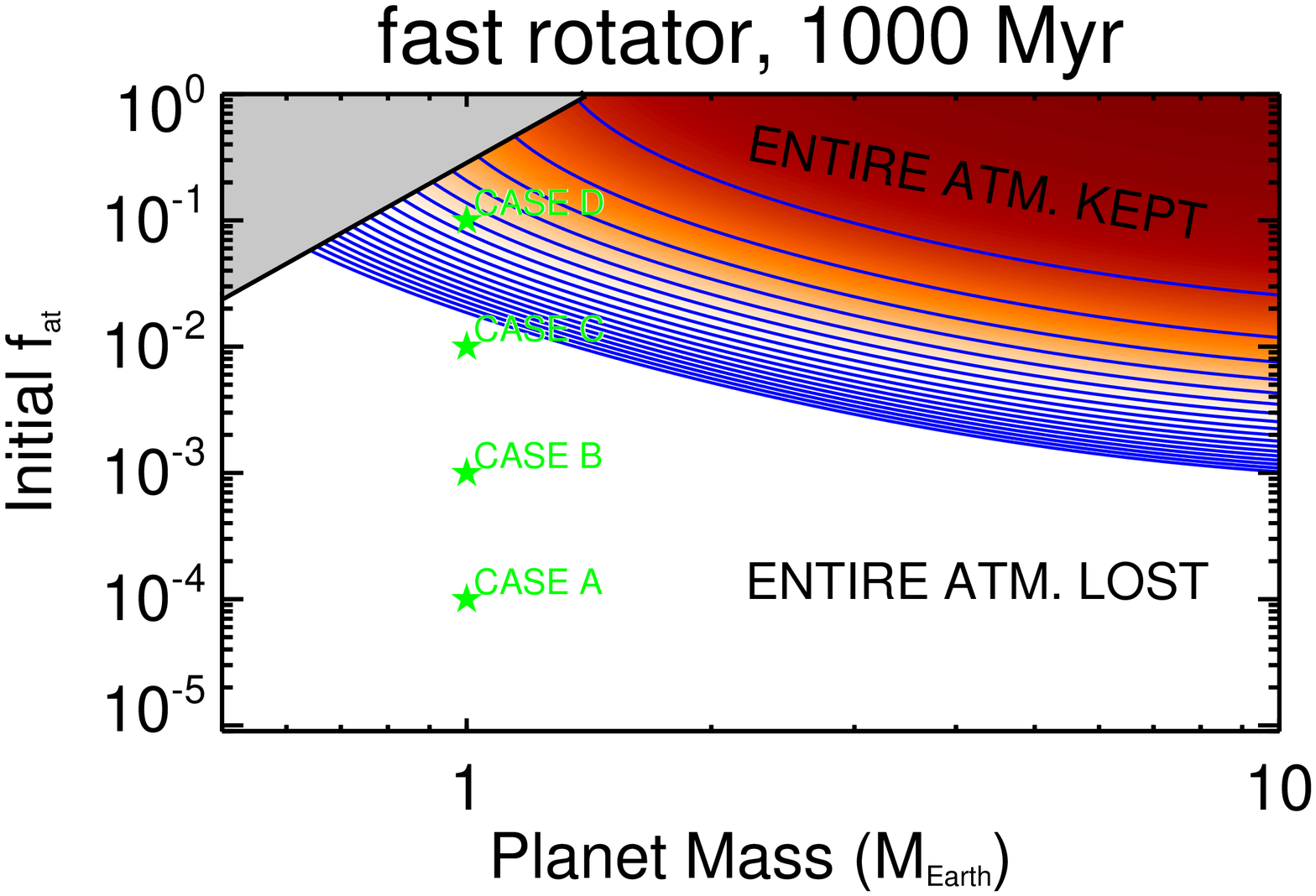}
\caption{
Contour plots showing the percentages of the initial hydrogen protoatmospheres remaining at 100~Myr (\emph{upper~row}) and 1000~Myr (\emph{lower~row}) as a function of planetary mass. 
In each panel, the contour lines show 20 evenly spaced levels, with dark red and white showing 100\% and 0\% of the initial atmospheric masses remaining.  
The left and right columns show the predictions assuming the star is a slow and a rapid rotator, corresponding to the red and blue lines in Fig.~\ref{fig:EarthCases}. 
The green symbols mark the cases shown in Fig.~\ref{fig:EarthCases}.
The grey shaded areas show the region where \mbox{$f_\text{at} > f_\text{at,max}$}, where $f_\text{at,max}$ is estimated from Eqn.~\ref{eqn:fatmax}.
} \label{fig:AllMasses}
\end{figure*}

\subsection{Stellar XUV evolution} \label{sect:TuTracks}

\citet{2015AA...577L...3T} developed a rotational evolution model for solar mass stars at the 10th, 50th, and 90th percentiles of the rotational distributions based on the models of \citet{2013AA...556A..36G} and \citet{2015AA...577A..28J}. 
\citet{2015AA...577L...3T} combined their rotation tracks with an empirical relation between rotation and $L_\text{X}$ (\mbox{5-100\AA}) derived by \citet{2011ApJ...743...48W}, and the conversion between $L_\text{X}$ and $L_\text{EUV}$ (\mbox{100-920\AA}) derived by \citet{2011AA...532A...6S}, to predict evolutionary tracks for $L_\text{X}$ and $L_\text{EUV}$. 
For the pre-main-sequence phase, they used a time dependent saturation threshold calculated using the stellar evolution models of \citet{2013ApJ...776...87S}. 
Their XUV evolutionary tracks are shown in Fig.~\ref{fig:EarthCases}.
Although we use these tracks in this letter, a set of simple power laws that approximately describe each track can also be found in \citet{2015AA...577L...3T}.

\section{Results: the importance of stellar rotational evolution} \label{sect:results}

By combing the XUV evolution tracks with Eqn.~\ref{eqn:Mdoteqn} and Eqn.~\ref{eqn:R0scaling}, we calculate the evolution of hydrogen atmospheres for planets with core masses between 0.5 and 10~M$_\oplus$, and the initial atmospheric masses between $10^{-5}$ and 1~M$_\text{pl}$. 
We integrate the atmospheric masses from 10~Myr to 1~Gyr taking into account the time variations in $\dot{M}_\text{at}$ due to both the stellar XUV evolution and the decreasing $R_0$ as a result of atmospheric erosion. 
We stop at 1~Gyr because at that age, $F_\text{XUV}$ drops well below 100~erg~s$^{-1}$~cm$^{-2}$.
This is problematic because at low $F_\text{XUV}$, the planetary wind at the exobase is slower than the escape velocity (see the middle-left panel of Fig.~\ref{fig:hydrosims}) meaning that many atoms in the flow should return to the planet after traveling on ballistic trajectories in the exosphere.

The cases in our grid break down into four interesting regimes illustrated in Fig.~\ref{fig:EarthCases} for Earth mass planets with different initial atmospheric masses. 
In Case~A, the initial atmospheric mass ($10^{-4}$~M$_\oplus$) is so small that everything is removed very quickly for all rotation tracks. 
In Case~D, the atmosphere ($10^{-1}$~M$_\oplus$) is so massive that thermal escape is negligible in all cases. 
In Case~B, the entire atmosphere ($10^{-3}$~M$_\oplus$) is removed in all cases, but the time required depends critically on the star's initial rotation rate; for the slow and fast rotators, the atmosphere is removed in 50~Myr and 250~Myr, respectively. 
The most dramatic difference between the rotation tracks is in Case~C, where the initial atmospheric mass ($10^{-2}$~M$_\oplus$) is such that if the planet is orbiting the rapid rotator, the atmosphere is removed in 400~Myr, whereas if the planet is orbiting the slow rotator, 70\% of the atmosphere remains at 1~Gyr. 
Due to the low stellar activity, the subsequent evolution of the atmosphere is negligible and the hydrogen envelope will never be lost.

In Fig.~\ref{fig:AllMasses}, we show how much atmosphere remains at ages of 100~Myr and 1~Gyr for all planetary masses. 
The boundary between the regimes where the planet has lost and retained its atmosphere is not only dependent on planetary mass and the age of the system, but also strongly dependent on the initial rotation rate of the host star. 
The cases shown in Fig.~\ref{fig:EarthCases} for Earth mass planets are similar for all planetary masses, with the only difference being that the boundaries between the regimes are shifted. 
The grey shaded areas show where our calculations are unrealistic because low mass planets will not collect and hold onto such massive atmospheres (\citealt{2014MNRAS.439.3225L}; \citealt{2015AA...576A..87S}).

\section{Discussion} \label{sect:discussion}

In this letter, we develop a method for estimating the evaporation of hydrogen atmospheres and give scaling laws that can easily be applied.
We show that the initial rotation rate of the central star can be fundamentally important for the evolution of a planetary atmosphere.
In all cases, we assume no intrinsic large scale planetary magnetic field, which could influence the flow and change $\dot{M}_\text{pl}$ if the gas is highly ionised (\citealt{2015ApJ...813...50K}). 
We have also only considered thermal mass loss due to the heating of the upper atmosphere. 
For hydrogen atmospheres, thermal mass loss dominates (\citealt{2013AsBio..13.1030K}; \citealt{2014AA...562A.116K}), and this is also likely to be the case for atmospheres made of heavier species if the XUV energy input is high enough (\citealt{2008JGRE..113.5008T}; \citealt{2009GeoRL..36.2205T}).  
For example, \citet{2015EPSL.432..126T} showed that thermal escape of O$_2$ from initially H$_2$O dominated atmospheres can be significant in certain cases.
In other cases, the mass loss might be dominated by non-thermal processes such as stellar wind charge exchange and pickup (\citealt{2010Icar..210....1L}).
The erosion of non-hydrogen dominated atmospheres therefore depends strongly on stellar winds; easily applicable models for the properties and evolution of stellar winds were given by \citet{2015AA...577A..28J} and \citet{2015AA...577A..27J}.

Our results provide further confirmation of the conclusions of \citet{2014MNRAS.439.3225L} that higher mass habitable-zone terrestrial planets could have difficulty losing hydrogen envelopes if they form in the circumstellar disk.  
Lower mass terrestrial planets, on the other hand, will lose their hydrogen atmospheres much more effectively.
For example, our results indicate that no \mbox{$\sim0.5M_\oplus$} planets will keep hydrogen atmospheres for more than 1~Gyr regardless of the activity evolutions of their host stars; this suggests that low-density habitable zone planets with such masses will only be found in young systems.

Although we have only considered hydrogen dominated protoatmospheres, the evolution of stellar rotation is also fundamentally important for the formation and evolution of secondary atmospheres and the development of habitability. 
For a planet to be habitable, it must first lose its protoatmosphere, and then it must also develop and retain an appropriate secondary atmosphere.  
Consider a habitable-zone Earth mass planet formed early enough to pick up a substantial hydrogen envelope. 
The initial rotation rate of the host star not only determines if the planet will lose its protoatmosphere, but also how long this process with take. 
A rapid rotator might cause this to happen very quickly, whereas a slow rotator might allow the atmosphere to remain for hundreds of Myr, potentially slowing down the solidification of the planet's surface and disrupting the formation of the secondary atmosphere.

Stellar magnetic activity evolves in non-trivial ways that depend sensitively on the star's rotational evolution.
Simple evolutionary decay laws that give one-to-one relations between age and XUV emission or winds are inappropriate at young ages. 
The aim of this letter, however, is not simply to study the importance of the initial stellar rotation rate on atmospheric evaporation; our results clearly demonstrate that a strong understanding of stellar activity evolution is an essential ingredient in detailed studies of the evolution of planetary atmospheres.

\acknowledgements{
The authors thank the anonymous referee for providing useful suggestions on our manuscript. 
This study was carried out with the support by the FWF NFN project S11601-N16 ``Pathways to Habitability: From Disk  to Active Stars, Planets and Life'' and the related subprojects S11602-N16, S11604-N16, and S11607-N16. 
LT was supported by an ``Emerging Fields" grant of the University of Vienna through the Faculty of Earth Sciences, Geography and Astronomy.
PO and HL acknowledge the FWF project P27256-N27.
}

\end{document}